\documentclass[twocolumn, twocolappendix]{aastex631}

\usepackage{comment}
\usepackage{hyperref}

\def\asec{$^{\prime\prime}$}

\definecolor{indigo}{rgb}{0.0, 0.25, 0.42}
\definecolor{forestgreen}{rgb}{0.13, 0.55, 0.13}
\definecolor{rtblue}{rgb}{0.25, 0.41, 0.88}

\graphicspath{{./}{figures/}}

\begin{document}

\title{JWST imaging of edge-on protoplanetary disks\\III.
Drastic morphological transformation across the mid-infrared in Oph163131}

\email{marion.villenave@unimi.it}
\author[0000-0002-8962-448X]{Marion Villenave}
\affiliation{Universit\'a degli Studi di Milano, Dipartimento di Fisica, via Celoria 16, 20133 Milano, Italy}
\affiliation{Jet Propulsion Laboratory, California Institute of Technology, 4800 Oak Grove Drive, Pasadena, CA 91109, USA}

\author[0000-0002-2805-7338]{Karl R. Stapelfeldt}
\affiliation{Jet Propulsion Laboratory, California Institute of Technology, 4800 Oak Grove Drive, Pasadena, CA 91109, USA}

\author[0000-0002-2805-7338]{Gaspard Duch\^ene}
\affiliation{Astronomy Department, University of California, Berkeley, CA 94720, USA}
\affiliation{Univ. Grenoble Alpes, CNRS, IPAG, F-38000 Grenoble, France}

\author[0000-0002-1637-7393]{Fran\c{c}ois M\'enard}
\affiliation{Univ. Grenoble Alpes, CNRS, IPAG, F-38000 Grenoble, France}

\author[0000-0002-3191-8151]{Marshall D. Perrin}
\affiliation{Space Telescope Science Institute, Baltimore, MD 21218, USA}

\author[0000-0001-5907-5179]{Christophe Pinte}
\affiliation{School of Physics and Astronomy, Monash University, Clayton Vic 3800, Australia}
\affiliation{Univ. Grenoble Alpes, CNRS, IPAG, F-38000 Grenoble, France}

\author[0000-0002-9977-8255]{Schuyler G. Wolff}
\affiliation{Department of Astronomy and Steward Observatory, University of Arizona, Tucson, AZ 85721, USA}

\author[0000-0003-1451-6836]{Ryo Tazaki}
\affiliation{Univ. Grenoble Alpes, CNRS, IPAG, F-38000 Grenoble, France}

\author[0000-0001-5334-5107]{Deborah L. Padgett}
\affiliation{Jet Propulsion Laboratory, California Institute of Technology, 4800 Oak Grove Drive, Pasadena, CA 91109, USA}

\begin{abstract}

We present JWST broadband images of the highly inclined protoplanetary disk SSTc2d J163131.2-242627 (Oph163131) from 2.0 to 21$\mu$m.  The images show a remarkable evolution in disk structure with wavelength, quite different from previous JWST observations of other edge-on disks. At 2.0 and 4.4$\mu$m, Oph163131 shows two scattering surfaces separated by a dark lane, typical of highly inclined disks. Starting at 7.7$\mu$m however, 1) the two linear nebulosities flanking the dark lane disappear; 2) the brighter nebula tracing the disk upper surface transitions into a compact central source distinctly larger than the JWST PSF and whose intrinsic size increases with wavelength; and 3) patches of extended emission appear at low latitudes, and at surprisingly large radii nearly twice that of the scattered light seen with $HST$ and NIRCam, and of the gas. We interpret the compact central source as thermal emission from the star and the inner disk that is not seen directly, but which instead is able to progressively propagate to greater distances at longer wavelengths. The lack of sharp-edged structures in the extended patchy emission argues against the presence of shocks and suggests photoexcitation or stochastic heating of material smoothly flowing away from the star along the disk surface.  Finally, the dark lane thickness decreases significantly between 0.6$\mu$m and 4.4$\mu$m which indicates that the surface layers of Oph163131 lack  grains larger than 1$\mu$m.  

\end{abstract}
\keywords{{Protoplanetary disks (1300); Planet formation (1241); Radiative transfer (1335); Dust continuum emission (412)}}

\section{Introduction}

Planets are built from the growth of sub-micron sized particles that evolve to larger bodies within protoplanetary disks.
A full understanding of this process requires a detailed characterization of the radial and vertical structure of disks. Indeed, the level of dust concentration within a disk plays an important role on the local grain growth~\citep[e.g.,][]{Birnstiel_2012} and planetesimal formation efficiency~\citep[e.g.,][]{Youdin_2007, Lambrechts_2012}. 
Pressure maxima can trap and concentrate dust radially, while size-dependent vertical settling concentrates dust vertically. 
These mechanisms can provide favorable conditions for rapid grain growth but remain to be fully characterized.

Protoplanetary disks viewed edge-on provide a unique opportunity to study their vertical structure, since it is directly visible.
Previous observations with the Hubble Space Telescope ($HST$) revealed that, in the optical and near-infrared, edge-on protoplanetary disks appear as two reflection nebulae separated by a dark lane~\citep[e.g.,][]{Burrows_1996, Watson_2007}. 
When observed at multiple scattered light wavelengths, the thickness of the dark lane can provide insight into the dust properties and disk vertical structure. Pioneering studies in three edge-on disks observed between 0.6$\mu$m and 5$\mu$m~\citep[HH~30, HV Tau C, HK Tau B;][]{cotera_2001, Watson_2004, Duchene_2010, McCabe_2011}  
showed that the dark lane thickness tends to decrease with wavelength. This indicates either a vertically uniform distribution of grain sizes whose dust opacity strongly decreases with wavelength,  
or a vertically stratified dust distribution with the larger grains settled nearer to the midplane.  

The James Webb Space Telescope (JWST), with its much higher sensitivity in the mid-infrared, allows such studies to be extended to more systems. JWST also brings a larger range of wavelengths in reach, by allowing the detection scattered light out to $\sim$20$\mu$m, which is extremely difficult from the ground. 
This is the third paper based on an ongoing JWST near- to mid-infrared imaging campaign targeting some of the largest edge-on disks in nearby star-forming regions (GO programs 2562 and 4290 in Cycles 1 and 2, co-PIs: F. M\'enard and K. R. Stapelfeldt). The goal of this program is to investigate dust evolution mechanisms such as dust vertical settling and grain growth, in particular through the wavelength evolution of their dark lane thickness. The first two papers of this series focused on the Class~II disk Tau042021~\citep{Duchene_2023}, and the Class~I disk IRAS04302~\citep{Villenave_2023b}.  In both systems, the dark lane thickness does not vary significantly with wavelength across the mid-infrared, which indicates that vertical settling is not taking place for grains $\lesssim$ 10$\mu$m in size. The last target of the JWST cycle 1 program, the Class II HH 30, will be analysed in Tazaki et al. (in prep). 

In addition to allowing the study of dust vertical settling, the new JWST observations also revealed other unexpected and remarkable disk features. For example, a surprising switch in the brightest nebulae with wavelength in IRAS04302 suggested that this disk could possess a tilted inner region. Tilted inner region also predict lateral asymmetries in the scattering nebulae when observed with the right orientation. Because lateral asymmetries are seen in a large number of edge-on disks, \cite{Villenave_2023b} suggested that tilted inner disks might be common in protoplanetary disks. This is consistent with shadows identified in scattered light observations of lower inclination systems~\citep[e.g.,][for a review]{Benisty_2023}, and could be caused by misaligned companion~\citep[e.g.,][]{Facchini_2013} or late infall~\citep[e.g.,][]{Kuffmeier_2021}. 
Finally, in the case of Tau042021, outflow structures well away from the disk plane were seen.  The clear X-shape feature detected in the broadband observations~\citep{Duchene_2023} was also identified in H$_2$ and PAH lines~\citep{Arulanantham_2024} and has been interpreted as part of a disk wind.  

The focus of this work is the Class~II edge-on protoplanetary disk SSTc2dJ163131.2-242627 (hereafter Oph163131). The system was identified on the basis of its double-peaked spectral energy distribution in the $Spitzer$ c2d survey \citep[][]{Evans_2009}, and confirmed as an edge-on disk with $HST$ imaging by \citet[][]{Stapelfeldt_2014}. It has been the target of detailed studies at several wavelengths, going from the optical, near infrared to the millimeter, both in dust and gas. The millimeter dust of this disk is found to be significantly affected by vertical settling, indicating particularly weak turbulence~\citep{Villenave_2022, Wolff_2021}. 
The 2D temperature structure of the disk revealed that the cold disk midplane is flanked by warmer CO gas at the disk's upper and lower surfaces.  At large radii, however, the CO is found to be vertically isothermal, which might be related to external UV irradiation of the disk \citep[][]{Flores_2021}. Finally, contrary to IRAS04302 which has an envelope~\citep{Wolf_2003, Villenave_2023b}, or Tau042021 which has a clear jet and winds~\citep{Duchene_2023, Arulanantham_2024}, no spectroscopic signatures of accretion or outflow were identified in Oph163131~\citep{Flores_2021}, making this source akin to a weak-line T Tauri stars and thus possibly the most evolved disk target in our JWST cycle 1 program.

Here, we extend the study of this system to the mid-infrared, aiming to understand the spatial distribution of intermediate sized particles. We present the observations in Sect. \ref{sec:observations}. Descriptions of the images are shown in Sect. \ref{sec:results} and discussed along with illustrative models in Sect. \ref{sec:discussion}. Finally, we summarize our findings in Sect.~\ref{sec:conclusions}.

\section{Observations \& Data reduction}
\label{sec:observations}
We observed Oph163131 with JWST (GO program 2562, PIs F.M\'enard \& K. Stapelfeldt) using both NIRCam and MIRI instruments in two consecutive visits starting on 2023 March 07 UT 19:56. The observations used 5 different filters, F200W ($2.0\mu$m), F444W ($4.44\mu$m), F770W ($7.7\mu$m), F1280W ($12.8\mu$m), and F2100W  ($21.0\mu$m). The two NIRCam images were obtained simultaneously with exposures of 773s. The MIRI observations were taken successively with exposures of 840s at 7.7$\mu$m, 855s at 12.8$\mu$m and 661s at 21$\mu$m, respectively. 

\begin{figure*}
    \centering
    \includegraphics[width = \textwidth]{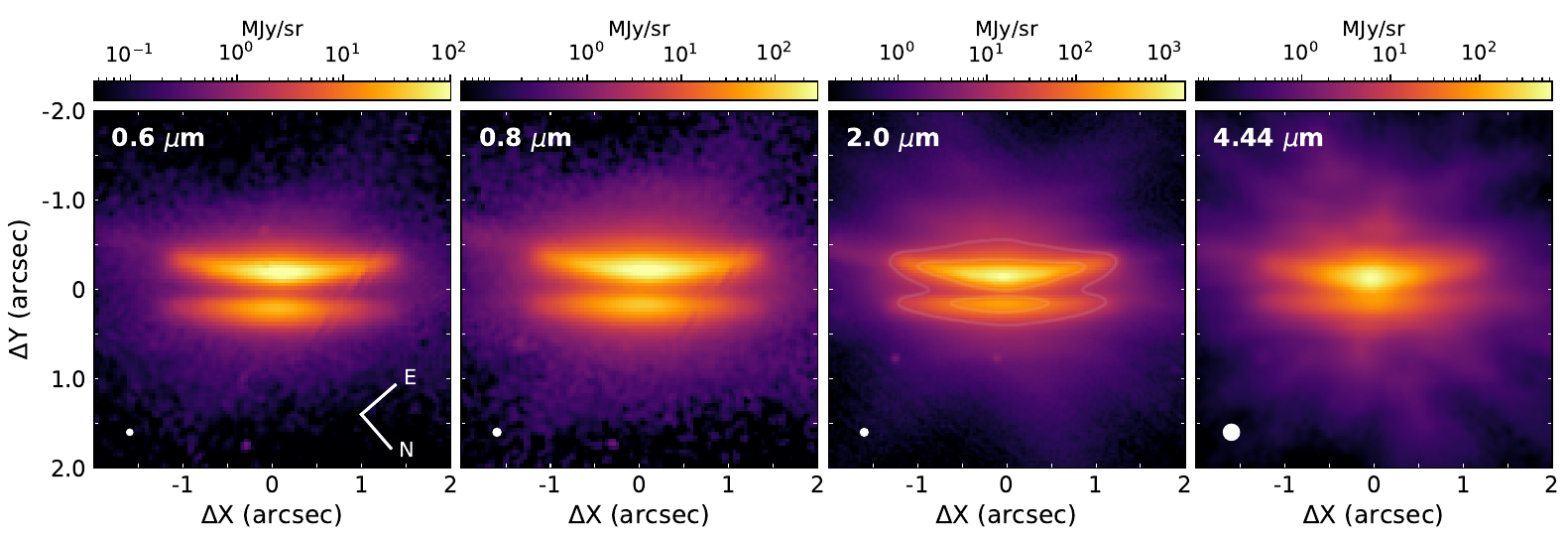}
    \caption{HST/ACS (0.6$\mu$m, 0.8$\mu$m) and JWST/NIRCam (2.0$\mu$m, 4.44$\mu$m) images of Oph163131, rotated so that the brighter South-East nebula is up. The white ellipse in the bottom left of each panel represents the resolution of the observations. On the third panel, the faint white contours correspond to 15 MJy/sr and 100 MJy/sr levels in the 2.0$\mu$m NIRCam image.}
    \label{fig:nircam_hst}
\end{figure*} 

\begin{figure*}
    \centering
    \includegraphics[width = 0.9\textwidth]{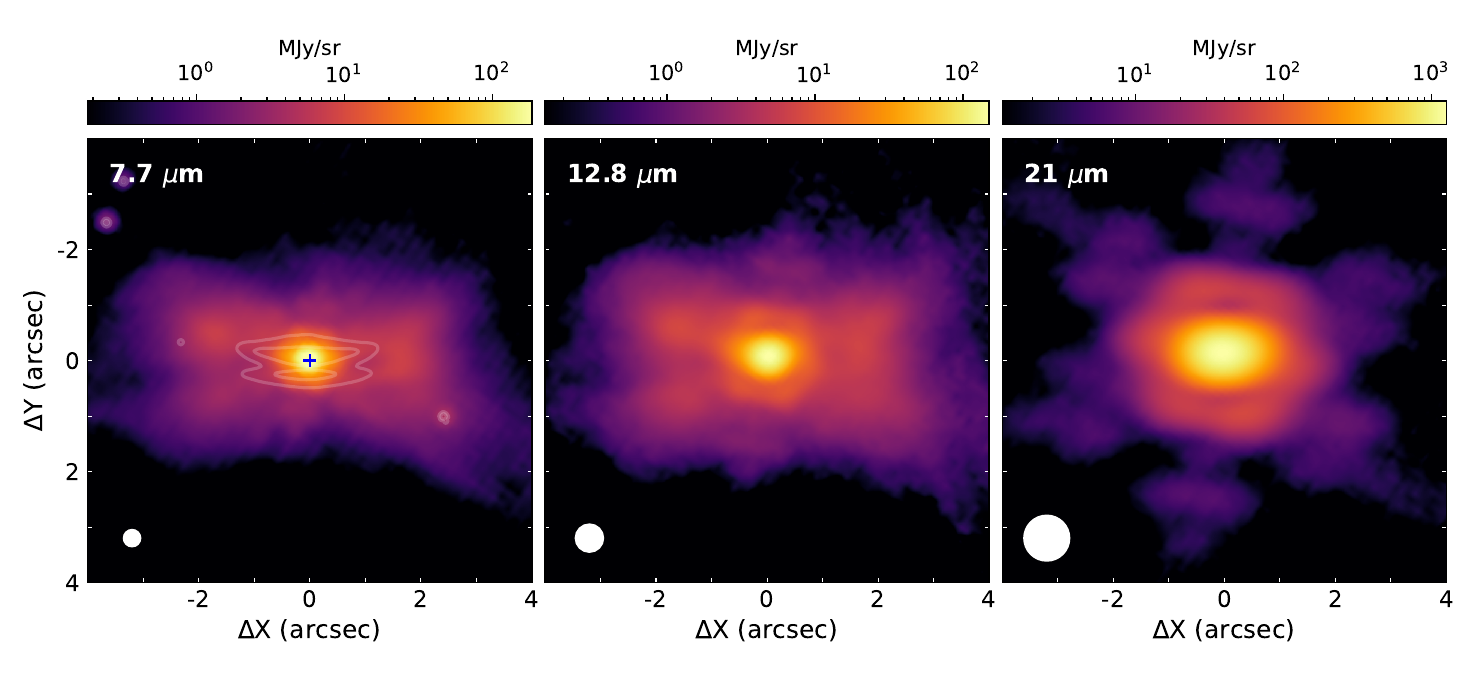}
    \vspace{-0.5cm}
    \caption{JWST/MIRI images of Oph163131. While the orientation is the same as in Fig. \ref{fig:nircam_hst}, the field of view is twice as large.  On the leftmost panel, the white contours correspond to 15 MJy/sr and 100 MJy/sr levels in the 2.0$\mu$m NIRCam image (same as in Fig.~\ref{fig:nircam_hst}), and the dark cross shows the location of the brightest pixel at 7.7$\mu$m. The resolution of the observations is shown as white ellipses in the bottom left corner. }
    \label{fig:miri}
\end{figure*}

\begin{figure}
    \centering
    \includegraphics[width =0.45\textwidth]{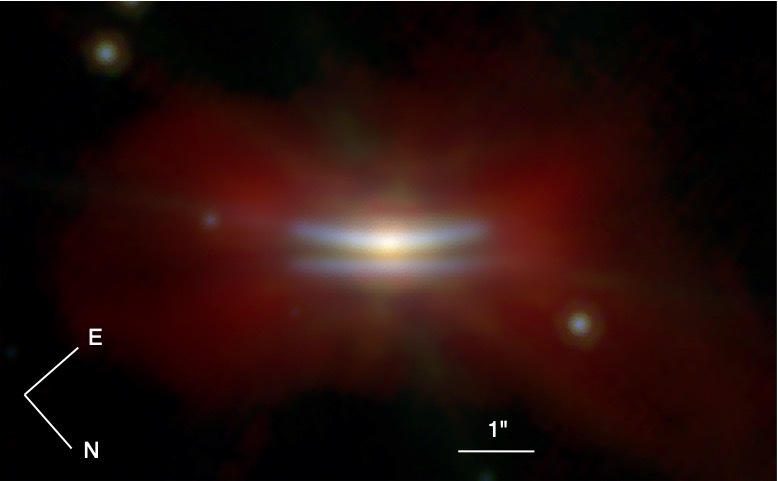}
    \caption{JWST 3-color composite image of Oph163131;  the full field of
view spans  10$\times$6\arcsec. The 2.0, 4.44, and 7.7 $\mu$m images are rendered (using a logarithmic stretch) in the blue, green, and red channels, respectively. }
    \label{fig:3color}
\end{figure}

For the F200W, F770W, F1280W, and F2100W observations, we obtained the phase 2 pipeline calibrated data from the MAST archive (DOI: \dataset[10.17909/zrsp-nc48]{https://doi.org/10.17909/zrsp-nc48}), and we re-ran phase 3, using the JWST pipeline version 1.9.5. In the \texttt{skymatch} step, we used the default skymethod \texttt{`global+match'} but set \texttt{subtract = True}.  
The subtracted background levels were 0.49, 15.76, 48.63, and 271.98 MJy/sr in F200W, F770W, F1280W, and F2100W, respectively. 
For the MIRI images, we also performed an additional 2D background subtraction step using the \texttt{Background2D} function in the \texttt{photutils} python package.

For the F444W observations, we obtained the uncalibrated data from the archive and re-ran each pipeline step using pipeline version  1.11.1.dev16+gb79a88af. In order to recover saturated pixels, we set \texttt{suppress\_one\_group = False} in the \texttt{ramp fit} step of phase 1. This allows us to suppress the ramp fit in the case when only the 0$^{th}$ group is unsaturated.  We use the default pipeline values in phase 2 and 3, except for the \texttt{skymatch} step where we subtract the background. The  subtracted background level in F444W was 0.65~MJy/sr.

Finally, we resampled and aligned each image to the NIRCAM F200W pixel size and position. We adopt the \texttt{reproject} python package to resample the images, and align them using a background galaxy visible in all fields and previously anchored to its position, as given in the GAIA DR3 catalog~\citep{Gaia_DR3} accessed through the Aladin sky atlas~\citep{Bonnarel_2000, Baumann_2022}. 
We visually checked that the alignment does not require a rotation; it only includes a shift in right ascension and declination. The final images are presented in Fig.~\ref{fig:nircam_hst}, Fig.~\ref{fig:miri}, and a 3-color composite image is displayed in Fig.~\ref{fig:3color}. \\ 

In this work, we also use the ACS/WFC images from the $HST$ Archive (program 12514, PI: K. Stapelfeldt). The $HST$ observations were obtained using filters F606W and F814W ($0.6\mu$m, $0.8\mu$m) and observed on March 4, 2012. They were previously published by \citet{Wolff_2021}. 
 
\section{Results}
\label{sec:results}

\begin{table*} 
    \caption{Oph163131 morphological properties}
    \centering
    \begin{tabular}{lccccccccc}
    \hline\hline
    $\lambda$ & $d_{neb}$ & FR$_{integrated}$&   $R_{FWHM}$ &$R_{FWHM}$ & $R_{FW10\%}$ & $R_{FW10\%}$\\ 
    ($\mu$m)& ('') & T/B  & B  ('') & T  ('') &B  ('') &T ('')\\
    \hline
    0.6 &  $ 0.42 \pm 0.01 $ & 3.0&  $ 0.93 \pm 0.11$ & $ 0.82 \pm 0.11 $ &  $1.88 \pm 0.08$ & $ 1.86 \pm 0.05 $  \\ 
    0.8    & $0.40 \pm 0.01$ &  3.3 & $0.99 \pm 0.12$  & $ 0.92 \pm 0.12$  & $2.00 \pm 0.04$  & $1.94 \pm 0.04$    \\  
    2.00   & $  0.34 \pm 0.01 $ &  3.5 & $ 1.12 \pm 0.10 $ & $ 0.76 \pm 0.07 $ &  $ 2.17 \pm 0.03 $ & $ 1.83 \pm 0.07 $  \\ 
    4.44   & $0.21 \pm 0.01$  & 3.8 &  $ 0.71\pm 0.07$  & $0.62 \pm 0.12$  & $1.84\pm 0.07$  & $1.33 \pm 0.08$  \\ 
    \hline
    \end{tabular} 
    \tablecomments{The dark lane thickness ($d_{neb}$), radial full width half maximum ($R_{FWHM}$), and the radial full width at 10\% of the peak ($R_{10\%}$) are inferred from fitting polynomial spines to the two scattering surfaces. The uncertainties correspond to the statistical error for the different spine averaging. The FR$_{integrated}$ column instead indicates to the integrated fluxe ratio between both nebula obtained using aperture photometry. We refer to the top (resp. bottom) nebula as ``T" (resp. ``B"). }
    \label{tab:darklane}
\end{table*}

In Fig.~\ref{fig:nircam_hst}, we show the NIRCam and HST observations of Oph163131 ($0.6 - 4.44\mu$m). At these wavelengths, the disk shows two scattered light nebulae separated by a darklane, typical of edge-on protoplanetary disks. We adopt the methodology previously employed by \cite{Duchene_2023} and \cite{Villenave_2023b} to fit the two scattering surfaces and quantify disk parameters. In short, this method finds a ``spine" for the two disk scattering surfaces. To do so, it first derives the vertical position of the scattering surfaces as a function of radius (i.e., distance from symmetry axis) defined as the peaks in cuts along the minor axis direction. Those are then fitted by a polynomial function to define the spines. 
We refer to these works for the full description of the methodology. We infer the darklane thickness ($d_{neb}$), radial full width half maximum ($R_{FWHM}$), and the radial full width at 10\% of the peak ($R_{10\%}$) of each spine. In addition, we also estimate the integrated flux ratio between the top and bottom nebulae, using aperture photometry. The results are summarized in Table~\ref{tab:darklane}. 

We find that the dark lane thickness $d_{neb}$ of Oph163131 decreases by a factor of two between 0.6$\mu$m and 4.44$\mu$m. 
While the radial extent of the top and bottom nebula are similar within 3$\sigma$, both become more centrally peaked at 4.44$\mu$m. The integrated flux ratio between the top and bottom nebulae increases with wavelength. \\

We show the observations obtained with MIRI, from 7.7$\mu$m to 21$\mu$m, in Fig \ref{fig:miri}. The morphology of the source is dramatically different at these wavelengths. Fig.~\ref{fig:3color} shows a 3-color composite image highlighting the major changes in the appearance of the source between NIRCam and MIRI wavelengths. 
At the longer wavelengths, the two disk scattering surfaces become invisible, and instead the images display a bright central source resolved along the disk major axis. We measured the size of this central source by fitting a 2D Gaussian using the IDP3 routines in IDL, comparing the observed source size to $WebbPSF$ models \citep{Perrin_2014} for a point source in each of the three MIRI filters.  After subtracting in quadrature the telescope beam size, also assumed to be Gaussian, 
we find that the intrinsic size of the central source is distinctly larger than the telescope beam size and increases with wavelength along the disk major axis (see Table \ref{tab:central_source}). 
This is a unique behavior among edge-on disks, where in some cases the central star peers through the disk in the near-infrared but appears as an unresolved source.

In addition, at 7.7$\mu$m and 12.8$\mu$m, the images show spatially extended diffuse emission extending up to radii twice that of the scattered light seen with HST and NIRCam.  
We assessed whether an optical problem or systematic issue might contribute to the extended appearance of the system at 7.7$\mu$m.  A JWST optical problem can be ruled out as the origin of this for several reasons. First and most simply, the numerous background stars observed in the field surrounding Oph163131 have normal, sharply focused point spread function (PSF) morphology in all filters. Secondly, the JWST wavefront sensing measurements taken on the days before and after these science observations confirm the telescope mirrors were in excellent alignment during this time period (65~nm rms  telescope wavefront error in the measurement closest in time to these observations).  Thirdly, the extended nebula seen at 7.7$\mu$m and 12.8$\mu$m is \textit{not} aligned with JWST's PSF hexagonal diffraction spikes, nor with MIRI's detector short-wavelength cruciform artifact. 
Thus we can rule out that any optical issue caused the extended diffuse/fluffy appearance in these 7.7$\mu$m and 12.8$\mu$m MIRI data; the observed morphology has to be astrophysical.\\

Finally, we use aperture photometry to determine the flux of the source at the different wavelengths. For all filters, we use  a large aperture ($7\farcs5\times4\farcs5$) that encompasses all the emission from the source. In addition, for MIRI wavelengths, we also consider a smaller aperture ($2"\times2"$), to illustrate the contribution from the central source. The results are reported in Table \ref{tab:photometry}. The fluxes are in general agreement with the values reported in \cite{Wolff_2021}, but at 7.7$\mu$m and 12.8$\mu$m our JWST fluxes are slightly higher than previous estimates from Spitzer and WISE.

\begin{table}[]
    \centering
    \caption{FWHM of the central source at MIRI wavelengths, along the disk major axis }
    \begin{tabular}{crrr}
    \hline
        $\lambda$ & Observed & $WebbPSF$ & Deconvolved \\
        ($\mu$m) & (\asec) & model (\asec) & (\asec), (au)\\
        \hline
         7.7 &  0.44 & 0.24 & 0.37, 54 \\
        12.8 &  0.65 & 0.45 & 0.47, 70 \\
        21.0 &  1.09 & 0.71 & 0.83, 122 \\
        \hline
    \end{tabular}
    \label{tab:central_source}
\end{table} 

\begin{table}[]
    \centering
    \caption{Photometry of Oph163131 }
    \begin{tabular}{ccc}
    \hline
        $\lambda$ & $F_{7\farcs5\times4\farcs5}$ & $F_{2''\times2''}$\\
        ($\mu$m) & (mJy) & (mJy)\\
        \hline
        2.0 &  $6.5 \pm 0.2$& $-$\\
        4.44 & $2.5 \pm 0.1$& $-$ \\
        7.70 & $2.9 \pm 0.1$& $1.6 \pm 0.1$\\
        12.8 & $3.2 \pm 0.1$& $1.9 \pm 0.1$ \\
        21.0 & $35.2 \pm 1.1$& $28.5 \pm 0.9$\\
        \hline
    \end{tabular}
    \label{tab:photometry}
     \tablecomments{We assume absolute flux calibration uncertainties of $\sim$3\% based on the JWST user documentation.}
\end{table}

\section{Discussion}
\label{sec:discussion}
\subsection{Chromaticity of the Dark Lane from 0.6-4.4 $\mu$m}

In Sect \ref{sec:results}, we showed that the dark lane thickness decreases by a factor 2 between 0.6$\mu$m and 4.4$\mu$m in Oph163131 (see Table~\ref{tab:darklane}). This result contrasts significantly with previous JWST observations of edge-on disks showing minimal variation in darklane thickness between the near- and mid-infrared \citep[$2-21 \mu$m; Tau042021, IRAS04302, L1527;][]{Duchene_2023, Villenave_2023b}. However, other optical to near infrared scattered light observations have also identified disks with large darklane thickness variations between wavelengths \citep[e.g., HK Tau B, HV Tau C, summarized in Fig. 3 of][]{Duchene_2023, McCabe_2011, Duchene_2010}. This suggests diversity in the structure and grain properties in protoplanetary disks, which we aim to investigate further in this section.\\
 
To interpret the wavelength variation of the dark lane thickness in Oph163131, we produce a suite of radiative transfer models with different properties, following the approach of \cite{Duchene_2023}. 
The main model parameters correspond to those of model A of \citet{Wolff_2021}, which reproduce well the scattered light (HST) observations of the disk. 
That model includes 
an inclination of 84.5$^\circ$, a gas scale height of 7.2au at a radius of 100au, a flaring exponent of 1.5, and a surface density exponent of 1.01. 
Following \citet{Duchene_2023}, we adopt the DIANA dust composition~\citep{Woitke_2016} and use the Distribution of Hollow Spheres method~\citep{Min_2016} to mimic compact dust aggregates. We also note that the high flaring exponent assumed here (based on previous parametric modeling of the source) should not significantly affect our qualitative analysis, as it provided the correct vertical extent at the disk outer edge, which is of interest here, in combination with the other disk parameters such as the surface density profile.  

\begin{table*}
\begin{center}
\caption{Summary of Model Exploration \label{tab:model_res}}
\begin{tabular}{ccccccccc}
\hline
Model & \multicolumn{3}{c}{Grain Size Distribution} & \multicolumn{2}{c}{Settling} & Total & Dark Lane \\
  & $p$ & $a_\mathrm{min}$ & $a_\mathrm{max}$&$\eta_\mathrm{settl}$ & $a_\mathrm{mix}$ & Dust Mass & Thickness\\
 &  & ($\mu$m)& ($\mu$m) &  & ($\mu$m) & ($10^{-4}\,M_\odot$) &  \\
\hline
Standard & 3.5 & 0.03 & 1000& \multicolumn{2}{c}{None} & 3.94 & $\times$\\
Dust Settling (10) & 3.5 & 0.03 & 1000& 0.5 & 10 & 4.59 & $\times$ \\
Dust Settling (1) & 3.5 & 0.03 & 1000& 0.5 & 1 & 7.10 & $\times$ \\
Dust Settling (0.1) & 3.5 & 0.03 & 1000& 0.5 & 0.1 & 1400 & (\checkmark)  \\
Pristine Dust  & 3.0 & {0.03} & {0.35}& \multicolumn{2}{c}{None} & 1.94 &  (\checkmark) \\
&  {1.5} & {0.35} & {100}& &&\\
Only small grains & 3.5 & 0.03 & 0.5& \multicolumn{2}{c}{None} & 0.43 &  (\checkmark) \\
\hline
\end{tabular}
\end{center}
\tablecomments{The symbols in the last column describe the degree to which a given model matches observations: a (\checkmark) indicates a qualitatively better match, while a $\times$ symbol points to a large shortcoming of the model.}
\end{table*}

\begin{figure}
    \centering
    \includegraphics[width =0.48\textwidth, trim = {0.5cm 0 1cm 1cm}, clip]{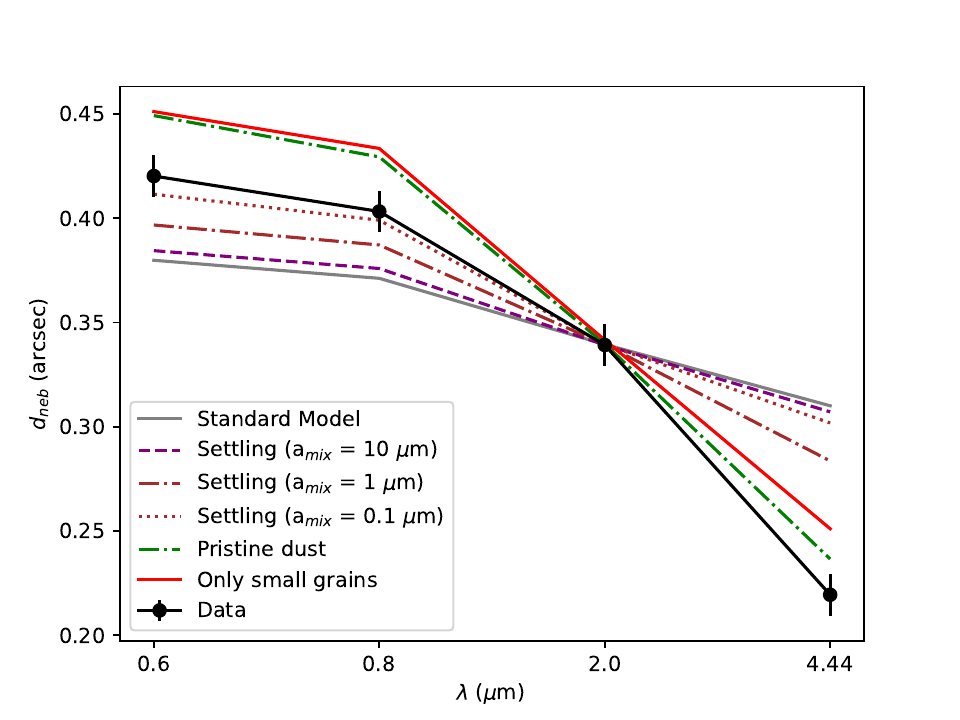}
    \caption{ Comparison of data and radiative transfer models of the dark lane thickness as a function of wavelength. The models with grains of 1$\mu$m or more present in the upper layers do not show sufficient variation with wavelength to reproduce the observations. } 

    \label{fig:model_chromaticity}
\end{figure}

We produce 6 different models which are distributed in three main families (``standard'', ``settled'', ``only small grains''). The standard model includes well mixed grains, with sizes following a power-law such that $n(a)da\propto a^{-p}da$, with $p=3.5$ and a maximum grain size of 1mm. Three models include parametric dust vertical settling, such that grains with size $a>a_{mix}$ have a reduced scale height of $h(a)\propto (a/a_{mix})^{\eta_{settl}}$, with $\eta_{settl} = 0.5$. 
We implement three scenarios, where grains larger than $a_{mix}=10\mu$m, 1$\mu$m, and 0.1$\mu$m are affected by vertical settling. 
Finally, we produce two models in the ``only small grain'' category. In the first one, we implement a pristine dust model with similar properties to that of \citet[][see their Table 3 and Appendix]{Duchene_2023}.This model includes two power-laws for the grain sizes. Grains of sizes ranging from 0.03 to 0.35$\mu$m, follow a power-law exponent of $p=3$, while grains from 0.35 to 100$\mu$m have a power-law exponent of $p=1.5$. The power-laws are normalized such that 80\% of the total mass is in the large grains
component. The second model of this group consists of well mixed grains, with a maximum size of 0.5$\mu$m. A summary of model parameters is presented in Table~\ref{tab:model_res}. All models are convolved by a JWST PSF obtained from the $WebbPSF$ software~\citep{Perrin_2014}, and the dust mass of each model is adjusted so that its dark lane thickness at 2$\mu$m matches that of the observations. The comparison of the models and the data are shown in Fig~\ref{fig:model_chromaticity}. \\

As previously shown by \cite{Duchene_2023}, we find that if grains of 10$\mu$m (or more) are present in the upper layers of the disk, the models show only modest chromaticity. These models do not match the observations of Oph163131. Moreover, even the models where grains larger than 1$\mu$m are affected by settling do not show sufficient chromaticity to reproduce the steep decrease in dark lane thickness between 0.6$\mu$m and 4.4$\mu$m.

On the other hand, the model with pristine dust, the model including no grains larger than 0.5$\mu$m, or the model where grains larger than 0.1$\mu$m are affected by settling reproduce better the observations. They show a steeper curve between 0.8$\mu$m and 4.4$\mu$m. This exploration suggests that the surface layers of Oph163131 are largely depleted of grains larger than 1$\mu$m.

Yet, we note that none of the models perfectly fit the evolution of the dark lane thickness with wavelength. Settling of grains larger than 1$\mu$m needs to be extremely efficient in the disk to reproduce the observations, but also, a few of those grains might still be present in the upper layers, because neither the pristine dust nor the small grain only model perfectly match the observed chomaticity. This could suggest that the settling prescription used here is too simplistic. 

Finally, we also found that none of the models reproduce the lateral intensity profile seen in the data, namely that the top nebula becomes more centrally peaked at 4.4$\mu$m than at shorter wavelengths (Fig.~\ref{fig:nircam_hst}). The model where grains larger than 0.1$\mu$m are settled partially fits the chromaticity, but in this case, we found that the lateral intensity profile is much broader than what is observed. In this model, the surface is mostly dominated by grains smaller than the observed wavelength, and consequently, each grain shows nearly isotropic scattering with a bluish extinction opacity (Rayleigh-like scattering). To explain the observations, dust particles that exhibit forward scattering and a bluish dust opacity simultaneously would be needed. We speculate that fluffy dust aggregates consisting of small monomer grains might help, as these particles appear to have the required optical properties \citep{Tazaki_2019}, but more detailed studies need to be performed to draw a robust conclusion.\\ 

The modeling of Oph163131 indicates that 1$\mu$m grains are mostly absent from the upper layers of the disk, and are at most only located below its $\tau=1$ surface. In other words, this indicates that grains of 1$\mu$m and larger are decoupled from the gas in that region.  This implies that the ratio $\alpha/St$, where $\alpha$ represents the turbulence level and $St$ the Stokes number, is very low in the upper layers of the disk for these particles. 
Oph163131 is a well known disk in which extremely low level of midplane turbulence has been identified, based on the radiative transfer modeling of its millimeter emission~\citep[$h_{mm}/h_g<10$, $\alpha_{mid}\leq10^{-5}$;][]{Villenave_2022}. The low midplane turbulence level of Oph163131 and the absence of micron sized particles in its upper layers suggest that the overall vertical turbulence $\alpha$ of this disk is particularly low, even at high altitude above the midplane. Alternatively, the Stokes number of particles of 1$\mu$m and larger may be particularly high. This could be the case if the gas surface density is very low in the upper layers of the disk, or potentially if grains are porous, as such grains have larger cross sections for the same mass, therefore increasing their Stokes number.\\

This result differs from those of previous studies on the variation of dark lane thickness of younger edge-on disks using JWST observations. Indeed, those showed that 10$\mu$m grains can be present in the upper layers of disks, even in the Class 0 stage~\citep[][Tau042021, IRAS04302, L1527]{Duchene_2023, Villenave_2023b}, indicative of a higher $\alpha/St$ ratio in these disks than in Oph163131, for particles of similar sizes. 
All of these disks have also been studied at millimeter wavelengths, revealing different levels of vertical concentration of these large dust particles. The younger L1527 and IRAS04302 appear vertically thick and not affected by settling~\citep{vantHoff_2023, Lin_2023, Villenave_2023}. This suggests no significant decoupling of particles up to $\sim$1mm.  
At the other end, in the more evolved Class~II Tau042021, dust appears settled at millimeter wavelengths~\citep{Villenave_2020}. This indicates that large grains eventually decouple from the gas, but for significantly larger particles ($a\gtrsim10\mu$m) than in the case of Oph163131 ($a\lesssim1\mu$m), possibly due to either lower turbulence, lower gas surface density, or different grain properties in Oph163131. Further statistical studies of disks at different evolutionary stages are needed to test whether such characteristics are common or if Oph163131 is an exception.

\begin{figure*}
    \centering
    \includegraphics[width =\textwidth]{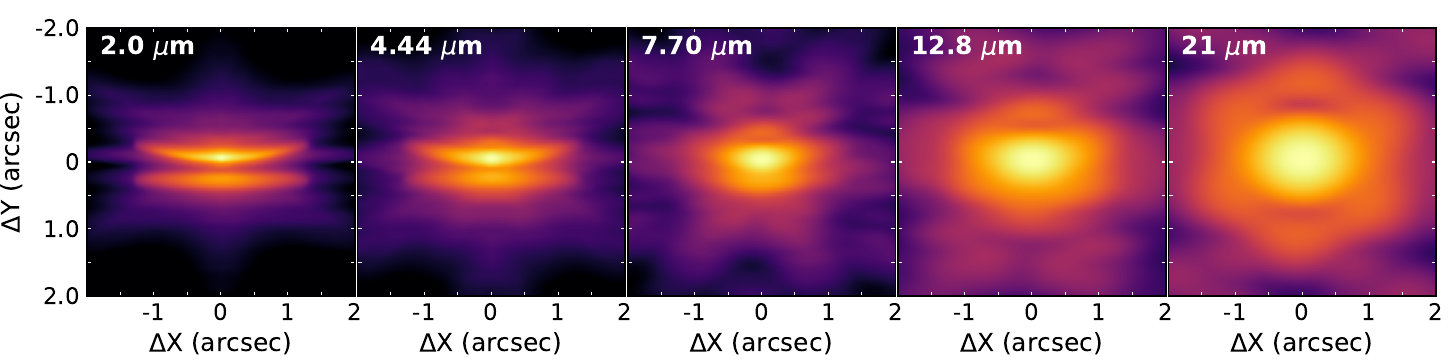}
    \caption{{Model A from \citet{Wolff_2021}, including only one smooth disk region, computed at  JWST wavelengths. Contrary to the data, this model does not become dominated by a central source starting at 7.7$\mu$m and does not include extended emission at 7.7$\mu$m and 12.8$\mu$m. The fields of view of the panels 7.7$\mu$m, 12.8$\mu$m, and 21$\mu$m are twice smaller than those in Fig~\ref{fig:miri}.}}
    \label{fig:modelA}
\end{figure*}

\begin{figure*}
    \centering
    \includegraphics[width =\textwidth]{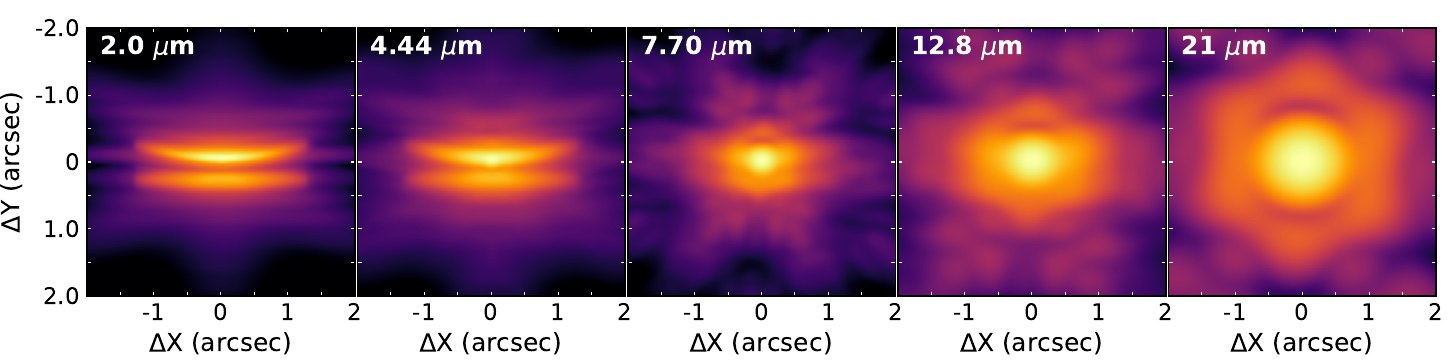}
    \caption{Toy model including a very massive and vertically thick inner region, allowing to reproduce a sharp transition in the source appearance beyond 4.4$\mu$m. The fields of view of the panels 7.7$\mu$m, 12.8$\mu$m, and 21$\mu$m are twice smaller than those in Fig~\ref{fig:miri}. The compact source appearing at 7.7$\mu$m and longer wavelengths is not centered on the position of the star in the model, but offset by 20mas or more toward the upper nebulae.} 
    \label{fig:model_transition}
\end{figure*}

\subsection{The Transition in Disk Appearance Beyond 4.4$\mu$m}
\label{sec:transition}

As shown in Sect. \ref{sec:results}, the disk appearance undergoes a sharp transition beyond 4.4$\mu$m. Beyond this wavelength, the two reflection nebulae are not clearly visible, and the disk appearance becomes dominated by a bright central source and extended fluffy emission.  
The central source at MIRI wavelengths is also centered onto the top scattered light nebula (see Fig.~\ref{fig:miri}). 

The appearance of a bright and compact central source is not expected in a smooth disk, even at the inclination of Oph163131 ($\sim85^\circ$). To illustrate this point, we computed the smooth model~A of \citet{Wolff_2021} at all JWST wavelengths and show the images in Fig.~\ref{fig:modelA}. The two scattered light surfaces are dominating the source appearance up to 21$\mu$m, even though they are vertically unresolved at wavelength larger than 12.8$\mu$m. This model does not become dominated by a bright central source starting at 7.7$\mu$m, contrary to the data.\\

To explain the significant change of morphology with wavelengths, several configurations are possible. For instance, a drastic change in the dust scattering properties, leading to almost exclusively forward scattering at 7.7$\mu$m, might lead to such an important change in the disk morphology. This might for example happen if the observations at 7.7$\mu$m probe  significantly larger grains than those seen at 4.4$\mu$m but it is difficult to reconcile with our current understanding of dust properties and dust settling. Alternatively, the inner disk could become suddenly very bright at 7.7$\mu$m and directly visible such that it dominates over the outer disk brightness. This can be expected if the inner disk is tilted above the outer disk or in the case of a very vertically and optically thick inner disk, which could be seen through the outer disk as the latter became optically thin. In this section, we produce a toy model using radiative transfer to test this second option. 

The goal of our toy model is to test whether a vertically and optically thick inner disk could produce the observed change in morphology, and specifically, the appearance of a compact source and the disappearance of the scattering nebulae starting from 7.7$\mu$m. 
We modified model A from \cite{Wolff_2021} such that the inner 10au contain as much mass as the outer 100au of the disk (M$_{dust}=5\ 10^{-4}$ M$_\sun$), and such that the scale height is $\sim4$ times larger within 10au than at larger radii. 
This transition radius roughly corresponds to the size of the very bright inner region detected in millimeter wavelengths ($\gtrsim$ 5 times brighter than the rest of the disk). In the modeling of the ALMA observations, \citet{Villenave_2022} obtained a surface density more than one order of magnitude higher in that inner region than in the rest of the disk, providing support to the toy model scenario presented here. The millimeter dust height in this region was however not constrained with ALMA. 
The model images are convolved by the JWST PSF and are shown in Fig.~\ref{fig:model_transition}.  

We find that this model is able to produce a sharp transition in the shape of the source between 4.4$\mu$m and 7.7$\mu$m. 
At 4.4$\mu$m, in addition to the two scattering nebulae, there is a hint of the diffraction from the PSF, suggesting that a bright source starts to shine through, which is also seen in the data.  
Then, starting at 7.7$\mu$m, the emission becomes dominated by the inner top scattering surface which starts to scatter significantly.  Indeed, at all wavelengths the brightest pixel in the model image is offset by more than 20mas from the position of the star, which shows that the bright central emission does not correspond to the star itself. Consistent with observations, the spectral energy distribution of the model shows a sharp rise beyond 10$\mu$m, while the main morphological changes occur between 4.44$\mu$m and 7.7$\mu$m, which also supports this interpretation.
 
Enhancing both the mass and scale height of the inner disk in the model creates a sharp transition at 10au, where photons from the inner disk must first scatter very high up above the midplane before being directed towards the outer disk. The line of sight from the observer to that location goes through relatively little column density in the outer disk, and that line of sight becomes optically thin between 4.4$\mu$m and 7.7$\mu$m, hence the behavior of the model images. 
In other words, the compact central source can be interpreted as emission from the star and the inner disk that is not seen directly, but which instead is able to progressively propagate through the disk to greater distances at longer wavelengths.

Finally, we note that the central source in the model is not resolved, nor does its size increase with wavelengths, as seen in the data. The apparent deconvolved size of the central source in the data, along the major axis of the disk, goes from 54au to 122au between 7.7$\mu$m and 21$\mu$m (Table~\ref{tab:central_source}), which roughly corresponds to the location of the two rings modeled in the ALMA continuum observations~\citep{Villenave_2022}. However, at these wavelengths, thermal emission is expected to arise from within 5au, which is significantly smaller than the observed sizes. This supports the interpretation that the compact central source seen at MIRI wavelengths does not correspond to direct thermal emission but instead to scattered photons that were thermally emitted and are able to propagate further in the disk. Further modeling would be required to quantitatively explore the conditions for such behavior.

\subsection{The Nature of the Patchy Extended Emission}

\begin{figure}
    \centering 
    \includegraphics[width =0.5\textwidth]{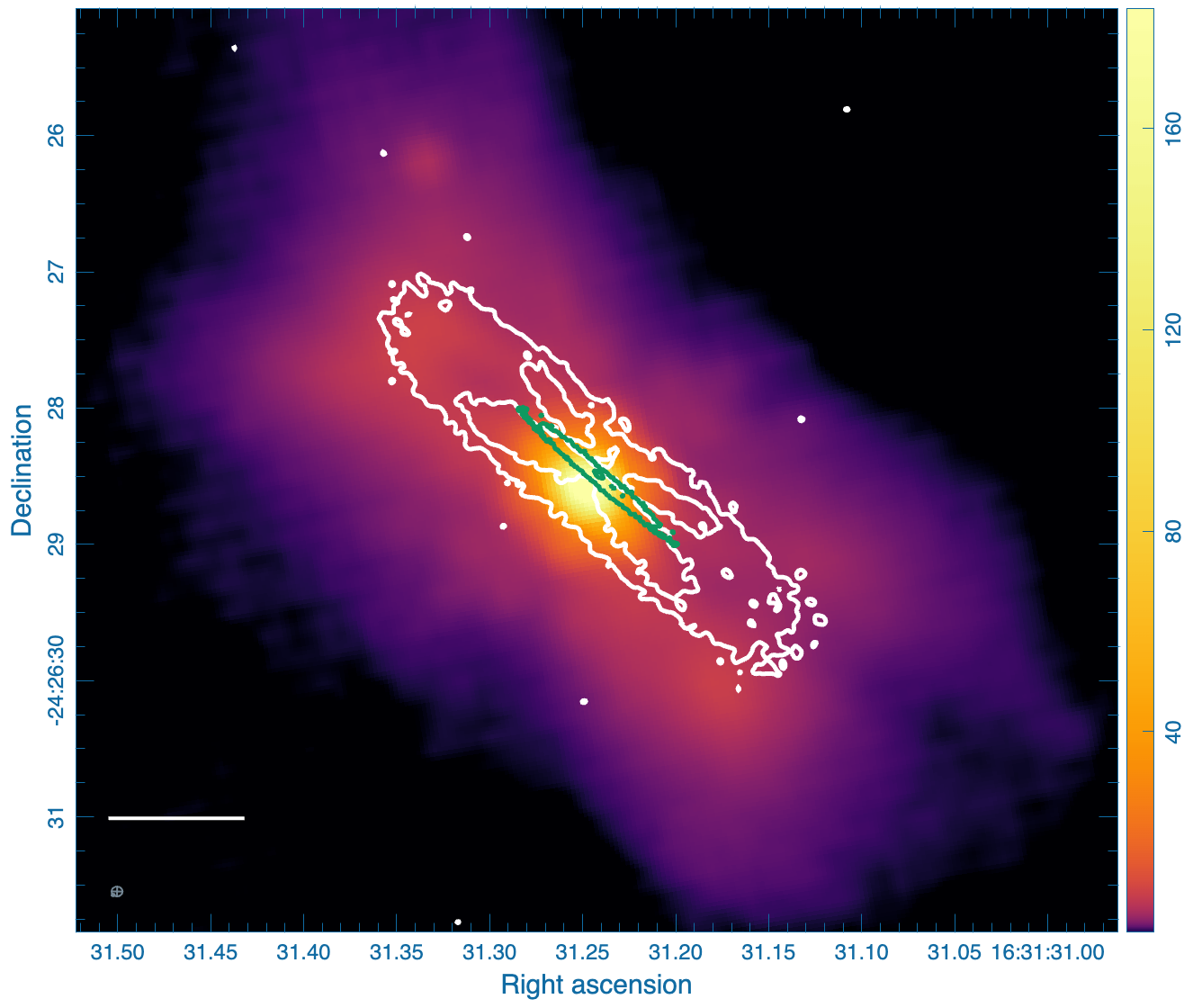}
    \caption{Overlay MIRI 7.7$\mu$m image, using the same colorscale as in Fig.~\ref{fig:miri}, and ALMA 1.3mm dust continuum (green contours, with levels 10$\sigma_{cont}$ and 20$\sigma_{cont}$) and $^{12}$CO observations (white contours, with levels 5$\sigma_{CO}$ and 20$\sigma_{CO}$), illustrating that the fluffy emission is located at higher altitude and larger radii than the gas emission. The ALMA data were published in \citet[][]{Villenave_2022}. A 1'' scale is represented by an horizontal line in the bottom left corner of the figure. } 
    \label{fig:almajwst}
\end{figure}

At 7.7$\mu$m and 12.8$\mu$m, the observed scattered light images show a significant contribution from extended emission (Fig.~\ref{fig:miri}). This feature does not appear axisymmetric  and potentially includes a dark, apparently shadowed region extending outward along the midplane, south-west and north-east of the source. In Fig.~\ref{fig:almajwst}, we overlay the MIRI 7.7$\mu$m image with millimeter continuum and $^{12}$CO gas emission probed by ALMA.  We find that the extended and faint emission is detected from the disk midplane to above the molecular disk as traced in CO. It is also extended to larger radii than the $^{12}$CO emission. The origin of this emission is currently unknown. It is however unlikely to be thermal continuum emission across the broad F770W and F1280W bandpasses, as any dust grains in the region would be expected to also produce scattered light signatures at other wavelengths, which are not observed.

Intriguingly, an extended feature had previously been detected in JWST images of the Class II edge-on disk Tau042021.  In that source, the extended feature was detected in the same JWST broadband filters as in Oph163131, centered on 7.7$\mu$m and 12.8$\mu$m, in the form of a sharp X-shape \citep{Duchene_2023}.  Using MIRI/MRS observations, \citet{Arulanantham_2024} revealed that the X-shape feature in Tau042021 corresponds to H$_2$ and PAH line emission. This could be indicative of a disk wind lifting material above the disk surface, and of shocks exciting the lines. 

In the case of Oph163131, the patchy emission is confined to a region much closer to the disk surface (i.e., with a larger opening angle compared to the minor axis direction than in Tau042021), and lacks sharp-edge features indicative of shocks. 
Photoexcitation or stochastic heating of material surrounding the source might explain the extended structure that we observe.  However, given the lack of spectral signatures of accretion in Oph163131~\citep{Flores_2021}, the flux of high energy photons from the central source should be much weaker than in Tau042021. Yet, the relatively symmetric appearance of the extended emission around the source and the suggestive disk shadow extending out into the patchy emission along the disk plane (Fig.~\ref{fig:miri}) suggests that the central source is illuminating the extended emission. 
Even though Oph163131 is not located in a PAH-bright region in the Spitzer IRAC Channel 4 maps, illumination by an external UV field, already suggested by \cite{Flores_2021} to explain the vertically isothermal profile in $^{12}$CO observations outside of 300au, is a possible alternative explanation.

\section{Conclusion}
\label{sec:conclusions}

We presented new JWST images of the Class II disk Oph163131. The two scattering surfaces of the disk are well visible up to 4.4$\mu$m. At longer wavelengths, the disk reveals a drastically different morphology.  A compact source suddenly appears between 4.4$\mu$m and 7.7$\mu$m, larger than the JWST PSF along the disk major axis, and increasing in size between 7.7$\mu$m and 21$\mu$m. Moreover, extended patchy emission is visible at 7.7$\mu$m and 12.8$\mu$m, extending twice as far as the scattered light disk seen at shorter wavelengths.

We find a significant decrease in dark lane thickness by a factor of $\sim2$ between 0.6$\mu$m and 4.4$\mu$m, which is consistent with no  grains larger than about $1\mu$m in the upper layers of the disk. For these particles, this implies that vertical settling dominates over dust stirring, keeping them below the $\tau=1$ surface. This could be explained by low turbulence up to high altitude above the disk midplane. 
This result however contrasts with constraints from younger sources with previous JWST observations, 
where larger grains of 10$\mu$m remained coupled to the gas and present in the disk upper layers. This points to a diversity in vertical settling/dust stirring efficiency between disks which needs to be be further explored.

The compact central source visible from 7.7$\mu$m is interpreted as emission from the star and inner disk that is not seen directly but scattered at progressively larger distances at longer wavelengths. 
We present a  toy model that can produce the sudden transition from parallel linear nebulae to a compact central source, which employs a very vertically and optically thick inner disk.
The model however does not  reproduce the size increase in the compact emission with wavelength.

The nature of the extended patchy emission is unclear, but the lack of sharp-edged structures argues against the presence of shocks and suggests instead photoexcitation or stochastic heating of material smoothly flowing away from the star along the disk surface.  Additional modeling and JWST IFU measurements will be needed to fully understand the wavelength evolution of this remarkable system.

\medskip

\emph{Acknowledgements:}
M.V. thanks Giovanni Rosotti for the useful discussions. A portion of this research was carried out at the Jet Propulsion Laboratory, California Institute of Technology, under a contract with the National Aeronautics and Space Administration (80NM0018D0004). The research of M.V. was conducted in a large fraction and supported by an appointment to the NASA Postdoctoral Program at the NASA Jet Propulsion Laboratory, administered by Oak Ridge Associated Universities under contract with NASA. 
This project has received funding from the European Research Council (ERC) under the European Union’s Horizon Europe research and innovation program (grant agreement No. 101053020, project Dust2Planets, PI F. Ménard \& grant agreement No. 101039651, project DiscEvol, PI G. Rosotti).  Views and opinions expressed are, however, those of the author(s) only and do not necessarily reflect those of the European Union or the European Research Council. Neither the European Union nor the granting authority can be held responsible for them.  M.V., K.R.S., G.D., and S.G.W. acknowledge funding support from JWST GO program \#2562 provided by NASA through a grant from the Space Telescope Science Institute, which is operated by the Association of Universities for Research in Astronomy, Incorporated, under NASA contract NAS5-26555. 
The JWST data presented in this article were obtained from the Mikulski Archive for Space Telescopes (MAST) at the Space Telescope Science Institute. The specific observations analyzed can be accessed at the following DOI \dataset[(10.17909/zrsp-nc48)]{https://doi.org/10.17909/zrsp-nc48}
This paper makes use of the following ALMA data: ADS/JAO.ALMA\#2018.1.00958.S and 2016.1.00771.S, which were observed with the band 6 receivers~\citep{Ediss_2004, Kerr_2014}.


\facilities{JWST, HST, ALMA}

\software{\texttt{mcfost}~\citep{Pinte_2006, Pinte_2009}, \texttt{Matplotlib}~\citep{Hunter_2007}, \texttt{Numpy}~\citep{Harris_2020}, \texttt{CARTA}~\citep{comrie_carta_2021}.}

\bibliography{biblio}{}
\bibliographystyle{aasjournal}

\end{document}